\documentclass[aps,prd,twocolumn,floatfix,nofootinbib,superscriptaddress]{revtex4-2}
\usepackage{dcolumn}

\usepackage{amsmath,amsfonts,extarrows,amssymb,mathrsfs,times,bm}
\usepackage{extarrows}
\usepackage{graphicx}
\usepackage{float}
\usepackage{graphicx,epstopdf}
\usepackage{dcolumn}
\usepackage{exscale}
\usepackage{relsize}
\usepackage{tikz}
\usepackage{mathtools}
\usepackage{mhchem}
\usepackage{amsthm}
\usetikzlibrary{positioning}
\usepackage[colorlinks=true,breaklinks=true,linkcolor=blue,citecolor=blue,urlcolor=blue]{hyperref}
\newcommand{\nn}{\nonumber}

\begin{document}
	
	\title{Generating New Spacetimes through Zermelo Navigation}
	\author{Zonghai Li}
	\affiliation{Center for Astrophysics, School of Physics and Technology, Wuhan University, Wuhan 430072, China}
	\date{\today}
	\author{Junji Jia}
	\email[Corresponding author:~]{junjijia@whu.edu.cn}
	\affiliation{Center for Astrophysics \& MOE Key Laboratory of Artificial Micro- and Nano-structures, School of Physics and Technology, Wuhan University, Wuhan, 430072, China}
	
	\begin{abstract}
		
	Zermelo navigation is not only a fundamental tool in Finsler geometry but also a fundamental approach to the geometrization of dynamics in physics. In this paper, we consider the Zermelo navigation problem on optical Riemannian space and, via Zermelo/Randers/spacetime triangle, explore the generation of new spacetimes from pre-existing ones. Whether the Randers metric has reversible geodesics corresponds to the presence of time-reversal symmetry in the generated spacetime. In cases where the Randers metric has reversible geodesics, we utilize a radial vector field to generate new static spacetimes from existing ones. For example, we can generate Schwarzschild, Rindler, de Sitter, and Schwarzschild-de Sitter spacetimes from flat spacetime. In fact, the Zermelo navigation method allows for the derivation of a variety of static spacetimes from flat spacetime. For multi-parameter spacetimes, they can be generated through various navigation paths. However, for some spacetimes, not all navigation paths may exist. In the second scenario, when the Randers metric does not have reversible geodesics, we employ a rotational vector field to transform non-flat static metrics into slowly rotating spacetimes. Alternatively, using a mixed vector field, we generate slowly rotating spacetimes starting from flat spacetime. We provide examples of generating Kerr spacetimes and Kerr-de Sitter spacetimes.

	\end{abstract}
	\maketitle

	\section{Introduction}
	
	In 1931, Ernst Zermelo posed and answered a fundamental question: When a ship is navigating on calm waters (Euclidean space) and suddenly encounters a gentle breeze (a vector field, referred to as wind), how can one find the optimal path to reach the destination in the shortest possible time~\cite{Zermelo1931}? This problem, known as the Zermelo navigation problem, is essentially a time-optimal control problem. In 2003, Shen addressed the general case of the Zermelo navigation problem, which involves finding the shortest-time path for an object in a Finsler space when it is subject to both internal and external forces~\cite{shenzhongmin2003}. In the specific case of navigating in a time-independent weak wind within a Riemannian space, Shen~\cite{shenzhongmin2003} demonstrated that the shortest-time trajectory corresponds precisely to the geodesic of a particular Finsler metric known as the Randers metric~\cite{Randers}. Conversely, given a Randers metric, one can formulate a corresponding Zermelo navigation problem~\cite{BRS2004}. In essence, there exists an equivalence relation between the Randers metric and the Zermelo navigation problem.
	
   Today, Zermelo navigation, as a methodology, has evolved into a pivotal tool for characterizing and classification of Finsler metrics~\cite{BRS2004}. Simultaneously, within the field of physics, Zermelo navigation continues to captivate the attention of researchers, finding applications in acoustics and optics~\cite{Gibbons&Warnick}, quantum control~\cite{Russell&Stepney2014,Russell&Stepney2015,Brody&Meier2015,BGM}, quantum mechanics~\cite{Qzermelo}, and in the domain of relativity~\cite{Caponio}.

   Closely related to this paper is the work by Gibbons et al.\cite{Gibbons-stm}, where they introduced the spacetime representation of Randers space (or equivalently, its Zermelo navigation problem). Specifically, a $n$-dimensional Randers metric can be seen as the optical metric of a ($n+1$)-dimensional stationary spacetime. In other words, the geodesics of the Randers metric can serve as null geodesics for a higher one-dimensional stationary spacetime. The Zermelo/Randers/spacetime triangle allows us to translate a problem from one language into any of the other two languages, often resulting in significant simplifications, as exemplified in Ref.\cite{Gibbons-stm}. In the Zermelo/Randers/spacetime triangle, there is a scenario not previously explored in the literature: when the Randers metric has reversible geodesics (meaning its geodesics remain geodesics when their orientation is reversed). In this scenario, the geodesics of the Randers metric are equivalent to the geodesics of a Riemannian metric, and the corresponding ($n+1$)-dimensional spacetime is static.

	Optical geometry (or optical space), defined by the optical metric, also known as Fermat geometry, was introduced by Weyl in 1917~\cite{Op0-Weyl}. Based on Fermat principle, the spatial part of null geodesic in $(n+1)$-dimensional spacetime is considered as the geodesic of the corresponding $n$-dimensional optical geometry. Thus, it allows researchers to study the propagation of light in space without the need to consider the time dimension. Optical geometry finds extensive applications, particularly in gravitational lensing and mathematical physics~\cite{GW2008,Werner2012,Op1-GW,Op2-PW,Op3-PW}. For a static spacetime, the optical metric is Riemannian, while for a stationary spacetime, it becomes a Finsler metric of the Randers type.
	
	For any ($n+1$)-dimensional static and spherically symmetric spacetime, we can consider a Zermelo navigation problem on its $n$-dimensional optical geometry. Solving this problem leads to an $n$-dimensional Randers space. We can describe this Randers space using an (n+1)-dimensional spacetime, which we conveniently refer to as the `Navigation Spacetime (NS)'. If the Randers metric has reversible geodesics, then the NS corresponds to a static and spherically symmetric spacetime (SSNS). Otherwise, the NS corresponds to stationary and axisymmetric spacetimes (SANS). On the other hand, adding parameters to a static spacetime leads to two outcomes: one preserves time-reversal symmetry, keeping the spacetime static, while the other breaks time-reversal symmetry, resulting in a stationary spacetime. For instance, adding an electric charge parameter transforms the Schwarzschild metric into the Reissner-Nordström (RN) metric, maintaining time-reversal symmetry. Conversely, adding an angular momentum parameter turns the Schwarzschild metric into the Kerr metric, breaking time-reversal symmetry.
	
\begin{widetext}
	\begin{figure}
		\centering
		\includegraphics[width=1\textwidth]{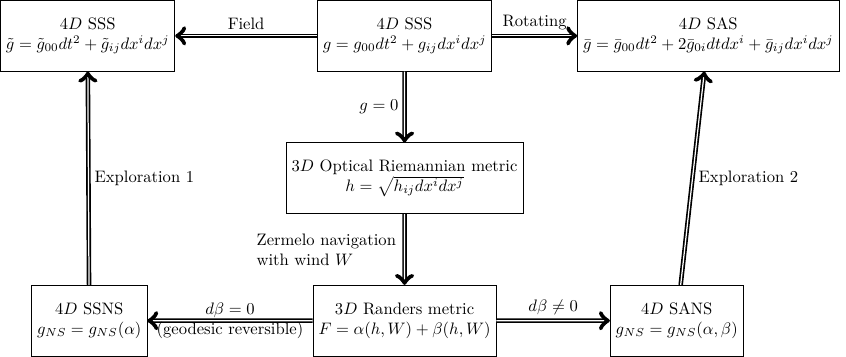}
		\caption{Summary of the research presented in this paper. }
		\label{figLI}
	\end{figure}
\end{widetext}

	Naturally, our interest lies in comprehending the relationship between NS and the physical spacetime obtained by adding parameters to the original static spacetime. If these two spacetimes are indeed related, it implies that we can derive a new spacetime from an existing one through navigation, without the need to solve field equations. The aim of this paper is to explore the generation of new spacetime from pre-existing ones using Zermelo navigation. We concentrate on $4D$ spacetime and investigate this problem in two scenarios based on whether the Randers metric has reversible geodesics. In the first scenario, when the Randers metric obtained through Zermelo navigation has reversible geodesics, new static spacetimes are generated. In the second scenario, where the Randers metric does not have reversible geodesics, we utilize navigation to generate rotating spacetimes from static spacetimes, focusing on the case of slow rotation. In both scenarios, we also consider the special case of generating new spacetimes from flat spacetime. The research conducted in this paper for these two scenarios is illustrated in Figure \ref{figLI}, and the symbols involved in this process will be introduced in subsequent sections.

	This paper is structured as follows. In Sec. \ref{ZRS-triangle}, we introduce Randers-Finsler metrics and provide an overview of the Zermelo/Randers/spacetime triangle. Sec. \ref{Zermeloopt} focuses on the Zermelo navigation problem in optical Riemannian space. In Sec. \ref{RVF}, we explore the generation of new static spacetimes from given static spacetimes using radial vector field navigation. In particular, we demonstrate that by commencing with flat spacetime (Minkowski spacetime) and applying navigation techniques, it is possible to derive various static spacetimes, including Schwarzschild spacetime, Rindler spacetime, de Sitter spacetime, and Schwarzschild-de Sitter (SdS) spacetime, among others. We also consider navigation based on other spacetimes. In Sec. \ref{RoVField}, we investigate the creation of slowly rotating spacetimes from given static spacetimes (excluding flat spacetime) using rotational vector field navigation. Alternatively, we employ a mixed vector field navigation to derive slowly rotating spacetimes starting from flat spacetime. We focus on two examples: the slowly rotating Kerr spacetime and the Kerr-de Sitter (KdS) spacetime. Finally, Sec. \ref{conclusion} provides a comprehensive summary and discussion of the entire manuscript. Throughout this paper, we use the convention $G = c = k_B = \hbar = 1$.

	\section{Zermelo/Randers/spacetime triangle}
	\label{ZRS-triangle}
	\subsection{Randers-Finsler metric}
Let $M$ be a smooth manifold of dimension $n$ with local coordinates $(x^i)$. Its tangent bundle is defined as $TM:=\bigcup_{x\in M}T_xM=\{(x,y)|x\in M,y=y^i\frac{\partial}{\partial x^i}\in T_xM\}$, where $T_xM$ is the tangent space at $x\in M$. A Finsler metric is a non-negative function defined on the tangent bundle $F: TM \to [0, \infty)$, satisfying the following properties~\cite{Cheng-shen}
	
	\begin{enumerate}
		\item $F$ is $C^{\infty}$ on $TM \setminus \{0\}$, where $\{0\}$ denotes the zero section of $TM$.
		
		\item $F(x, \lambda y) = \lambda F(x, y)$ for any $\lambda > 0$.
		
		\item The matrix composed of the fundamental tensor
		
		\begin{align}
			g_{ij}(x, y) = \frac{1}{2} \frac{\partial^2 F^2}{\partial y^i \partial y^j},
		\end{align}
		
		is positive-definite.
	\end{enumerate}
	
	The Riemannian metric $F = \sqrt{g_{ij}(x)y^iy^j}$ is a Finsler metric with a quadratic form restriction. One of the most common non-Riemannian Finsler metrics is the Randers metric defined as follows
	
	\begin{align}
		\label{Randersform}
		F= \alpha + \beta,
	\end{align}
	
	where $\alpha = \sqrt{\alpha_{ij}(x)y^iy^j}$ is a Riemannian metric, and $\beta = \beta_i(x)y^i$ is a 1-form on $M$, satisfying the strong convexity condition
	
	\begin{align}
		|\beta|^2 = \alpha^{ij}\beta_i\beta_j < 1.
	\end{align}
	
Different from Riemannian metrics, Finsler metrics admit non-reversibility or asymmetry, i.e., they allow for $F(x,-y)\neq F(x,y)$. The Randers metric~\eqref{Randersform} is reversible if and only if $\beta=0$, reducing it to a Riemannian metric. This non-reversibility in the metric leads to non-reversibility in geodesics and curvature~\cite{Shen-reversible}. A Finsler metric is said to have reversible geodesics if all of its geodesics remain geodesics when their orientation is reversed~\cite{Crampin-RR}. Interestingly, it is possible for a non-reversible Finsler metric to have reversible geodesics. For the Randers metric $F=\alpha+\beta$, Crampin~\cite{Crampin-RR} demonstrated that it has reversible geodesics if and only if $\beta$ is closed, i.e., $d\beta=0$.

 Additionally, the equation of motion for a free particle in Randers space is given by~\cite{Gibbons-stm}
	
	\begin{eqnarray}
		\label{motion-equation}
		\frac{Du^i}{dl} = (d\beta)^i_ju^j,
	\end{eqnarray}
	where $u^i = \frac{dx^i}{dl}$, and $l$ represents the arc length with respect to $\alpha$. When $d\beta=0$, it is evident that the geodesics of $F=\alpha+\beta$ coincide with the geodesics of $\alpha$. In other words, if $F$ has reversible geodesics, then its geodesics are identical to the geodesics of $\alpha$.

   On the other hand, it is evident from the equations of motion~\eqref{motion-equation} that it exhibits gauge invariance under the following transformation~\cite{Gibbons-stm}

\begin{align}
	\label{gauge-invariance}
	\beta\to \beta+d\Phi,
\end{align}
where $\Phi$ represents an arbitrary scalar field.

	\subsection{Zermelo navigation problem on Riemannian manifold}
	
   Assume a particle is moving in a Riemannian space $(M,h)$ and is subjected to the influence of a time-independent weak vector field (wind) $W$, satisfying $|W|^2=h_{ij}W^iW^j<1$. According to Shen~\cite{shenzhongmin2003}, the shortest-time trajectory of the particle corresponds to a geodesic of the following Randers metric
   \begin{equation}
   	\label{Randers-data}
	\begin{aligned}
		&F(x,y)=\sqrt{\alpha_{ij}y^iy^j}+\beta_iy^i,\\
		&\alpha_{ij}=\frac{\lambda h_{ij}+W_iW_j}{\lambda^2}, \\
		& \beta_i=-\frac{W_i}{\lambda},
	\end{aligned}
	\end{equation}
	where
	\begin{align}
		&&\lambda=1-h_{ij}W^i W^j,\quad W_i=h_{ij}W^j.\nn
	\end{align}
	
   Let $u$ represent the velocity generated by the internal force of $\alpha$, which corresponds to the speed of free particles in the absence of wind, satisfying $h_{ij}u^iu^j=1$. Wind can be considered as the velocity generated by an external force. The velocity of a particle under the combined influence of internal and external forces is $v=u+W$, which is the velocity generated by the internal force of $F=\alpha+\beta$. Therefore, Zermelo navigation provides a means to convert external forces into internal forces, thereby achieving a geometrization of dynamics. One crucial point to emphasize is that in the Zermelo navigation method, velocity or vector field plays a prominent role.

\subsection{The inverse problem of Zermelo navigation}
	
    For any Randers metric $(\alpha,\beta)$, can it be realized through the perturbation of some Riemannian metric $h$ by some vector field $W$ satisfying $h_{ij}W^i W^j<1$?
	
	The answer is yes. For a given Randers metric $(\alpha,\beta)$, there exists its corresponding Zermelo navigation problem $(h,W)$, as follows~\cite{BRS2004}
	
	\begin{equation}
		\label{Zermelo}
	\begin{aligned}
		&h_{ij}=\xi \left(\alpha_{ij}-\beta_i \beta_j\right), \\
	& W^i=-\frac{\beta^i}{\xi}.
	\end{aligned}
	\end{equation}
	where
	
	\begin{align}
		\label{Zermelo3}
		&\xi=1-\alpha^{ij}\beta_i \beta_j,\quad \beta^i=\alpha^{ij}\beta_j.\nn
	\end{align}
In conclusion, Randers data $(\alpha,\beta)$ and Zermelo navigation data $(h, W)$ are equivalent. It is worth noting that
	
	\begin{align}
		|W|^2=h_{ij}W^i W^j=\alpha^{ij}\beta_i \beta_j=|\beta|^2<1, \quad \lambda=\xi.
	\end{align}
This implies that the weakness of the wind corresponds to the strong convexity of the Randers metric.
	
	\subsection{Spacetime picture}
	According to Gibbons et al.~\cite{Gibbons-stm}, the geodesics of an $n$-dimensional Randers metric $(\alpha,\beta)$ can be interpreted as null geodesics in a family of $(n+1)$-dimensional stationary spacetimes given by
	
	\begin{align}
		\label{Randerspacetime}
		g_{R}=V^2\left[-(dt-\beta)^2+\alpha^2\right],
	\end{align}
	where $V^2$ represents the conformal factor, $\beta=\beta_idx^i$, and $\alpha^2=\alpha_{ij}dx^idx^j$.
	
When the Randers metric $F=\alpha+\beta$ has reversible geodesics (i.e., $d\beta=0$), as previously discussed, its geodesics coincide with those of $\alpha$. Therefore, we can neglect the $\beta$ term in the above spacetime metric, i.e., $g_R(\alpha,\beta)=g_R(\alpha)$, resulting in the static metric

\begin{align}
	\label{Randersspacetime-static}
	g_{R}=V^2\left[-dt^2+\alpha^2\right].
\end{align}

Using the transformation~\eqref{Randers-data}, the stationary spacetime metric~\eqref{Randerspacetime} in terms of Zermelo data $(h,W)$ can be written as

\begin{align}
	\label{spacetime-zermole}
	g_{Z}=V^2\left[-dt^2-\frac{2W_i}{\lambda}dtdx^i+\frac{h^2}{\lambda}\right],
\end{align}
where $h^2=h_{ij}dx^idx^j$ and $W_i=h_{ij}W^j$. Similarly, the static spacetime metric~\eqref{Randersspacetime-static} in terms of Zermelo data $(h,W)$ can be expressed as

\begin{align}
	\label{ZSSNS}
	g_{Z}=V^2\left(-dt^2+\frac{W_iW_j}{\lambda^2}dx^idx^j+\frac{h^2}{\lambda}\right).
\end{align}

In addition, any stationary spacetime can be represented using either Randers data $(\alpha, \beta)$ or Zermelo data $(h, W)$, as fallows~\cite{Gibbons-stm}
\begin{align}
	\label{Gibbons-triangle}
	g=&g_{00}dt^2+g_{0i}dtdx^i+g_{ij}dx^idx^j\nn\\
	=&g_{00}\left[-\left(dt-\beta_idx^i\right)^2+\alpha_{ij}dx^idx^j\right]\nn\\
	=&\frac{g_{00}}{1-h_{ij}dx^idx^j}\times\nn\\
	&\left[-dt^2+h_{ij}(dx^i-W^idt)(dx^j-W^jdt)\right].
\end{align}
Here, the Randers representation ($\alpha,\beta$) is the optical metric, while the Zermelo representation ($h,W$) corresponds to the spacetime in Painlevé-Gullstrand coordinates. Note that if the optical Randers metric $(\alpha,\beta)$ in Eq.\eqref{Gibbons-triangle} has reversible geodesics, then the stationary spacetime~\eqref{Gibbons-triangle} is indeed static.

	\section{Zermelo navigation on optical Riemannian space}
	\label{Zermeloopt}
    The metric for a $4D$ static spacetime in coordinates ($t,x^i$) ($i=1,2,3$) is given by
    \begin{align}
	\label{spacetimele}
	g=g_{00}dt^2+{g}_{ij}dx^idx^j.
    \end{align}

	By setting $g=0$, we can derive the optical metric, which is a $3D$ Riemannian metric, as follows
	
	\begin{align}
		\label{fensleranders}
		h=\sqrt{h_{ij}dx^idx^j}=\sqrt{-\frac{g_{ij}}{g_{00}}dx^idx^j},
	\end{align}
	where
	\begin{align}
		h_{ij}=-\frac{g_{ij}}{g_{00}}.
	\end{align}
	
	According to Fermat's principle, light rays are geodesics in space $(M^3,h)$, which is referred to as the optical space or optical geometry. Now, let us consider a Zermelo navigation problem in optical space. If we have a time-independent vector field $W=W(x)$ that satisfies $|W|^2=h_{ij}W^iW^j<1$, then according to Eq.~\eqref{Randers-data}, the solution corresponds to the following Randers metric:
	\begin{equation}
		\label{ORZnavo}
		\begin{aligned}
			&\alpha_{ij}=-\frac{g_{ij}}{g_{00}\lambda}+\frac{W_iW_j}{\lambda^2},\\
			& \beta_i=-\frac{W_i}{\lambda},
		\end{aligned}
	\end{equation}
	where
	\begin{align}
		\lambda=1+\frac{g_{ij}}{g_{00}}W^i W^j, \quad W_i=-\frac{g_{ij}}{g_{00}}W^j.\nn
	\end{align}

In Schwarzschild coordinates $(t, r, \theta, \phi)$, the SSS metric can be expressed as
	\begin{align}
		g = -A(r)dt^2+ B(r)dr^2 + C(r)d\Omega^2,
	\end{align}
where
	\begin{align}
		d\Omega^2= d\theta^2 + \sin^2\theta d\phi^2.
	\end{align}
	
The Zermelo data ($h,W$) becomes

\begin{equation}
	\left\{
	\begin{aligned}
		&h^2= \frac{1}{A^2}dr^2 + \frac{C}{A}d\Omega^2, \\
		&W = W(r,\theta)\frac{\partial}{\partial x^i},
	\end{aligned}
	\right.
\end{equation}
with $h_{ij}W^iW^j<1$.

The Rander metric~\eqref{ORZnavo} becomes

\begin{equation}
	\label{RandersSSS}
	\begin{aligned}
		&\alpha^2=\frac{B}{A\lambda}dr^2+\frac{C}{A\lambda}d\Omega^2+\frac{W_iW_j}{\lambda^2}dx^idx^j, \\
		& \beta=-\frac{W_i}{\lambda}dx^i,
	\end{aligned}
\end{equation}
where
\begin{align}
	\lambda=1-\frac{g_{ij}}{A}W^i W^j,\quad W_i=\frac{g_{ij}}{A}W^j.\nn
\end{align}

Now, by utilizing the Randers data $(\alpha,\beta)$ and the Zermelo data $(h,W)$, we can derive the expressions for the $4D$  spacetime~\eqref{Randerspacetime}-\eqref{ZSSNS}. In this context, we refer to the resulting spacetime as the navigation spacetime (NS), with the metric denoted as $g_{NS}=g_{R}=g_{Z}$.

	\section{Radial vector field and static metric}
	\label{RVF}
	\subsection{Radial vector field}
	We choose the following wind
	\begin{align}
		\label{R-wind}
		W=\mathcal{Q}(r)\frac{\partial}{\partial r},
	\end{align}
where the condition
\begin{align}
	\label{FCW}
	|W|^2 = h_{11}(W^1)^2 = \frac{B}{A}\mathcal{Q}^2 < 1,
\end{align}
is satisfied.
	
	Thus we have
	\begin{equation}
		\begin{aligned}
		&W_1=h_{11}W^1=\frac{B}{A}\mathcal{Q},\\
		&\lambda=1-|W|^2=1-\frac{B}{A}\mathcal{Q}^2.
	\end{aligned}
	\end{equation}

With these relations, the Randers metric \eqref{RandersSSS} can be written as follows

	\begin{equation}
		\label{RRdata}
			\begin{aligned}
			&\alpha^2=\frac{B}{A-B\mathcal{Q}^2}\left(1+\frac{B\mathcal{Q}^2}{A-B\mathcal{Q}^2}\right)dr^2\\
			&~~~~~~~~+\frac{C}{A-B\mathcal{Q}^2}d\Omega^2, \\
			& \beta=-\frac{B\mathcal{Q}}{A-B\mathcal{Q}^2}dr,
		\end{aligned}
	\end{equation}
	
Clearly, we have $d\beta=0$, which means that $\beta$ is closed. Equivalently, $F=\alpha+\beta$ has reversible geodesics. The geodesic of $F$ coincides with that of $\alpha$. From the perspective of gauge invariance~\eqref{gauge-invariance}, $\beta$ can be expressed as

\begin{align}
\beta=d\mathcal{Z}(r),
\end{align}
where

\begin{align}
\mathcal{Z}=-\int \frac{B(r)\mathcal{Q}(r)}{A(r)-B(r)\mathcal{Q}^2(r)}dr.
\end{align}
As a result, we can eliminate $\beta$, and all the navigation results are expressed in terms of the Riemannian metric $\alpha$.

Substituting the Randers metric given in Eq.\eqref{RRdata} into Eq.\eqref{Randersspacetime-static} and letting $V^2=A\lambda=A-B\mathcal{Q}^2$, the resulting NS metric $g_{NS}=g_{R}$ can be expressed as follows

\begin{align}
	\label{spscetimeR}
	g_{NS}=&-(A-B\mathcal{Q}^2)dt^2+\frac{AB}{A-B\mathcal{Q}^2}dr^2+Cd\Omega^2.
\end{align}

Note that Eq.~\eqref{FCW} implies that $A-B\mathcal{Q}^2>0$ ($A>0, B>0$). This means that if the NS metric describes the spacetime outside the black hole, then it only corresponds to the area outside the event horizon. The area where the navigation conditions fail is exactly the area where optical metrics cannot be defined.

The most interesting case is when applied to flat spacetime with $A=B=1$ and $C=r^2$,

	\begin{align}
		g=&-dt^2+dr^2+r^2d\Omega^2.
	\end{align}
	
 The corresponding optical metric is the Euclidean metric

	\begin{align}
		h^2=&dr^2+r^2d\Omega^2.
	\end{align}
	
	At this time, the NS metric~\eqref{spscetimeR} becomes
	
	\begin{align}
		\label{MinspscetimeR}
		g_{NS}=&-(1-\mathcal{Q}^2)dt^2+\frac{1}{1-\mathcal{Q}^2}dr^2+r^2d\Omega^2.
	\end{align}
	
	Therefore, any static spacetime of the form~\eqref{MinspscetimeR} can be obtained through Zermelo navigation on Euclidean space, with the following wind
	
	\begin{align}
		W=\pm \mathcal{Q}\frac{\partial}{\partial r},\quad |W|^2=Q^2<1.
	\end{align}

  Below, we provide some examples in a semi-Newtonian, semi-Einsteinian manner to illustrate this point.
	
	\subsection{Schwarzschild metric}
	
	Assuming that a particle with mass $m$ is only influenced by gravity in $3D$ Euclidean space, its speed $W$ generated by gravity can be determined as follows
	
	\begin{align}
		\label{NewtonTest}
		\frac{1}{2}m|W|^2=\frac{Mm}{r}.
	\end{align}
	
	Note that $|W|^2=h_{11}\mathcal{Q}^2=\mathcal{Q}^2$, we have $\mathcal{Q}^2=2M/r$. Now we consider the Zermelo pair
	
		\begin{equation}
		\label{Newton-wind}
		\left\{
		\begin{aligned}
			&h^2=dr^2+r^2d\Omega^2,\\
			&W=-\sqrt{\frac{2M}{r}}\frac{\partial}{\partial r},
		\end{aligned}
		\right.
	\end{equation}
	with
	
	\begin{align}
	|W|^2=\frac{2M}{r}<1.\nn
	\end{align}

Consequently, using $(h,W)$, the NS metric~\eqref{MinspscetimeR} becomes the Schwarzschild metric

		\begin{align}
			\label{Schwarzschild}
		g_{S}=&-\left(1-\frac{2M}{r}\right)dt^2+\frac{1}{1-\frac{2M}{r}}dr^2+r^2d\Omega^2,
	\end{align}
	
	If we consider the $\beta$ term and use $V^2=A-B\mathcal{Q}^2=1-\frac{2M}{r}$, the Zermelo data~\eqref{Newton-wind} leads to the NS metric~\eqref{ZSSNS} becoming
	
	\begin{align}
		\label{Schwarzschild-PG}
		g_{Spg}=&-\left(1-\frac{2M}{r}\right)dt^2+2\sqrt{\frac{2M}{r}}dtdr+dr^2+r^2d\Omega^2,\nn\\
		=&-dt^2+\left(dr+\sqrt{\frac{2M}{r}}dt\right)^2+r^2d\Omega^2,
	\end{align}
	which represents the Schwarzschild metric in Painlevé-Gullstrand coordinates. In accordance with Eq.~\eqref{Gibbons-triangle}, Gibbons et al. \cite{Gibbons-stm} derived the Zermelo pair~\eqref{Newton-wind} from the above equation.
	
	When confronted with the intricate representation of a metric in specific coordinates, we express it using the Randers representation, as demonstrated by the second equal sign in Eq.~\eqref{Gibbons-triangle}. If $d\beta=0$, we can subsequently eliminate $\beta$ from the metric, thereby exposing its true form.
	
	We navigate in optical space (in this example, $3D$ Euclidean space), where geodesics correspond to light rays. Therefore, in the process of deriving the wind as described above, we effectively assume that light possesses mass. We will continue with this assumption in the following discussion.
	
	\subsection{Rindler metric}
	
	In $3D$ Euclidean space, consider a particle with mass $m$ affected by a radial constant acceleration $\gamma$. The velocity field (W) generated by this acceleration can be obtained as follows
	
	\begin{align}
		\label{Rindler-acc}
		\frac{1}{2}m|W|^2=m\gamma r.
	\end{align}
	
	Consequently, we have $\mathcal{Q}^2=2\gamma r$. The Zermelo data becomes
	
		\begin{equation}
		\label{Rindler-wind}
		\left\{
		\begin{aligned}
			&h^2=dr^2+r^2d\Omega^2,\\
			&W=-\sqrt{2\gamma r}\frac{\partial}{\partial r},
		\end{aligned}
		\right.
	\end{equation}
Satisfying
	
	\begin{align}
		\label{MconditionR}
		|W|^2=2\gamma r<1.
	\end{align}
	
Thus, the NS spacetime~\eqref{spscetimeR} becomes the Rindler metric
	
	\begin{align}
		\label{Rindler}
		g_{Ri}=&-\left(1-2\gamma r\right)dt^2+\frac{1}{1-2\gamma r}dr^2+r^2d\Omega^2.
	\end{align}

The Rindler metric exhibits a horizon with a radius of $r_0=1/(2\gamma)$. The surface gravity corresponds to the acceleration $\gamma$, and the Hawking temperature is given by

\begin{align}
	T=&\frac{\gamma}{2\pi},
\end{align}
which is associated with the Unruh effect.

If we consider Cartesian coordinates $(t,x,y,z)$ and a constant acceleration $\gamma$ along the $x$ direction, then the NS metric becomes

	\begin{align}
		g_{Ri}=&-\left(1-2\gamma x\right)dt^2+\frac{1}{1-2\gamma x}dx^2+dy^2+dz^2.
	\end{align}
	
By introducing a new coordinate $X=\sqrt{\frac{1-\gamma^2x^2}{2\gamma}}$, the metric can be expressed as

	\begin{align}
		g_{Ri}=&-(\gamma X)^2dt^2+dX^2+dy^2+dz^2.
	\end{align}

\subsection{de Sitter metric}
	
In this context, we study the navigation problems associated with a positive cosmological constant $\Lambda$. Euclidean space endowed with a cosmological constant has an energy density given by

\begin{align}
	\rho=\frac{\Lambda}{8\pi}.
\end{align}

 Now, let us imagine a spherical region with a radius $R$. The mass enclosed within this region can be calculated as

	\begin{align}
		M(R)=\frac{4}{3}\pi R^3\rho=\frac{1}{6}R^3\Lambda.
	\end{align}
	
Suppose we have a particle with mass $m$ located on the surface of this spherical region, and it is influenced solely by the mass $M$. Using the equation

	\begin{align}
		\label{dSTest}
		\frac{1}{2}m|W|^2=\frac{M(R)m}{R}=\frac{1}{6}mR^2\Lambda,
	\end{align}
we can determine that $\mathcal{Q}^2=\frac{\Lambda}{3}R^2$. If we consider $R$ as a general radial coordinate $r$, we can derive the following Zermelo pair
	
	\begin{equation}
		\left\{
		\begin{aligned}
			&h^2=dr^2+r^2d\Omega^2,\\
			&W=-\sqrt {\frac{\Lambda}{3}}r,
		\end{aligned}
		\right.
	\end{equation}
with

	\begin{align}
		|W|^2=-\sqrt {\frac{\Lambda}{3}}R.
	\end{align}
	
Consequently, when we apply this Zermelo pair to the NS metric~\eqref{MinspscetimeR}, we obtain the de Sitter metric

	\begin{align}
		\label{dS-metric}
		g_{dS}=&-\left(1-\frac{\Lambda}{3}r^2\right)dt^2+\frac{1}{1-\frac{\Lambda}{3}r^2}dr^2+r^2d\Omega^2.
	\end{align}

\subsection{Schwarzschild-de Sitter metric}

To obtain the SdS metric using Zermelo navigation, three strategies can be employed. The first strategy, similar to the previous examples, starts with a flat metric and then derives the SdS metric by considering the combined effects of Newton wind and cosmological constant wind. The second strategy involves obtaining the SdS metric from the Schwarzschild metric and considering the cosmological constant wind. The third strategy involves obtaining the SdS metric from the de Sitter metric and considering the Newton wind. We will now consider these strategies individually.

\subsubsection{The navigation of $(h_M,W_{N+C})$}
Now, we consider the derivation of the SdS metric from a Minkowski metric.  Utilizing the equation

\begin{align}
	\frac{1}{2}m|W_{N+C}|^2=\frac{Mm}{r}+\frac{1}{6}mr^2\Lambda,
\end{align}
we find that $\mathcal{Q}_{N+C}^2=\frac{2M}{r}+\frac{\Lambda}{3}r^2$. Now, let us consider the Zermelo pair $(h_M,W_{N+C})$

\begin{equation}
	\left\{
		\begin{aligned}
			&h_M^2=dr^2+r^2d\Omega^2,\\
			&W_{N+C}=-\sqrt{\frac{2M}{r}+\frac{\Lambda}{3}r^2}~\frac{\partial}{\partial r},
		\end{aligned}
		\right.
	\end{equation}
	with
	
	\begin{align}
		|W_{N+C}|^2=\frac{2M}{r}+\frac{\Lambda}{3}r^2<1.\nn
	\end{align}
	
	The resulting NS metric~\eqref{MinspscetimeR} corresponds to the SdS metric
	
	\begin{align}
		\label{SdS-metric}
		g_{SdS}=&-\left(1-\frac{2M}{r}-\frac{\Lambda}{3}r^2\right)dt^2+r^2d\Omega^2\nn\\
		&+\left(1-\frac{2M}{r}-\frac{\Lambda}{3}r^2\right)^{-1}dr^2.
	\end{align}

\subsubsection{The navigation of $(h_S,W_{C})$}

For the Schwarzschild metric~\eqref{Schwarzschild}, its optical metric is given by

\begin{align}
	h_S^2=\frac{1}{\left(1-\frac{2M}{r}\right)^2}dr^2+\frac{r^2}{1-\frac{2M}{r}}d\Omega^2.
\end{align}

Consider a particle with mass $m$ moving in the optical space $(M^3,h_S)$, influenced by the cosmological constant $\gamma$. Using Eq.~\eqref{dSTest} and making the following substitutions

\begin{align}
	|W| \to|W_C|^2=&{(h_{S})}_{11}\mathcal{Q_{C}}^2\nn\\
	=&\frac{1}{(1-\frac{2M}{r})^2}\mathcal{Q_{C}}^2,\nn\\
	R\to\frac{r}{\sqrt{1-\frac{2M}{r}}}&,\nn
\end{align}
we obtain the equation

	\begin{align}
		\frac{1}{2}m\frac{1}{(1-\frac{2M}{r})^2}\mathcal{Q_C}^2=\frac{1}{6}m\frac{r^2}{1-\frac{2M}{r}}\Lambda,
	\end{align}
which leads to $\mathcal{Q_C}^2=\frac{\Lambda}{3}r^2(1-\frac{2M}{r})$

Consider the following Zermelo pair

\begin{equation}
	\left\{
		\begin{aligned}
			& h_S^2=\frac{1}{\left(1-\frac{2M}{r}\right)^2}dr^2+\frac{r^2}{1-\frac{2M}{r}}d\Omega^2,\\
			&W_C=-\sqrt{\frac{\Lambda}{3}r^2\left(1-\frac{2M}{r}\right)}\frac{\partial}{\partial r},\\
		\end{aligned}
		\right.
	\end{equation}
	with
	
	\begin{align}
		|W_C|^2=\frac{1}{1-\frac{2M}{r}}\frac{\Lambda}{3}r^2<1.
	\end{align}
	
	The NS metric~\eqref{spscetimeR} thus becomes the SdS metric, as given by Eq.~\eqref{SdS-metric}.

\subsubsection{The navigation of $(h_{dS},W_{N})$}

Now, let us explore how to obtain the SdS metric from the de Sitter metric~\eqref{dS-metric}. The optical metric for the de Sitter metric is given by

\begin{align}
	h_{dS}^2=\frac{1}{\left(1-\frac{\Lambda}{3}r^2\right)^2}dr^2+\frac{r^2}{1-\frac{\Lambda}{3}r^2}d\Omega^2.
\end{align}

Utilizing Eq.~\eqref{NewtonTest}, but with the following substitutions

\begin{align}
	&|W|^2\to|W_N|^2=\frac{1}{\left(1-\frac{\Lambda}{3}r^2\right)^2}\mathcal{Q}^2,\nn\\
	&M\to\frac{M}{(1-\frac{\Lambda}{3}r^2)^{3/2}},\nn\\
	&r\to \frac{r}{\sqrt{1-\frac{\Lambda}{3}r^2}},\nn
\end{align}
we derive

\begin{align}
	\frac{1}{2}m\frac{1}{\left(1-\frac{\Lambda}{3}r^2\right)^2}\mathcal{Q_N}^2=\frac{Mm}{r(1-\frac{\Lambda}{3}r^2)},
\end{align}
which leads to $\mathcal{Q_N}^2=\frac{2M}{r}(1-\frac{\Lambda}{3}r^2)$.

Now, consider the following Zermelo pair

\begin{equation}
	\left\{
		\begin{aligned}
			& h_{dS}^2=\frac{1}{\left(1-\frac{\Lambda}{3}r^2\right)^2}dr^2+\frac{r^2}{1-\frac{\Lambda}{3}r^2}d\Omega^2,\\
			&W_N=-\sqrt{\frac{2M}{r}\left(1-\frac{\Lambda}{3}r^2\right)}~\frac{\partial}{\partial r},
		\end{aligned}
		\right.
	\end{equation}
	with
	
	\begin{align}
		|W_N|^2=\frac{1}{1-\frac{\Lambda}{3}r^2}\frac{2M}{r}<1.
	\end{align}
	
The NS metric~\eqref{spscetimeR}, when employing Zermelo navigation with $(h_{dS},W_N)$, corresponds to the SdS metric~\eqref{SdS-metric}.

The example above illustrates that the same multi-parameter spacetime can be obtained from different spacetimes through navigation. However, for certain spacetimes, certain navigation paths do not exist. For instance, for the RN spacetime, it appears not to be obtainable from Schwarzschild spacetime through navigation in a real wind, although it can certainly be obtained from flat spacetime. Consider the following Zermelo pair

\begin{equation}
	\left\{
	\begin{aligned}
		&h^2=dr^2+r^2d\Omega^2,\\
		&W=-\sqrt{\frac{2M}{r}-\frac{Q^2}{r^2}}~\frac{\partial}{\partial r},
	\end{aligned}
	\right.
\end{equation}
requiring

\begin{align}
	|W|^2=\frac{2M}{r}-\frac{Q^2}{r^2}<1.
\end{align}
Then NS metric~\eqref{MinspscetimeR} leads to RN metric

\begin{align}
			\label{RNmetric}
			g_{RN}=&-\left(1-\frac{2M}{r}+\frac{Q^2}{r^2}\right)dt^2+\frac{1}{1-\frac{2M}{r}+\frac{Q^2}{r^2}}dr^2\nn\\
			&+r^2\left(d\theta^2+\sin^2\theta d\phi^2\right),
\end{align}
where $Q$ is the charge of the central black hole.

\section{Rotating vector field, mixed vector field and stationary metric}
	\label{RoVField}
	
\subsection{Rotating vector field}

We consider a vector field with only a non-zero $\phi$ component, and the navigation data is given by

\begin{equation}
	\label{ZP-rotating}
	\left\{
\begin{aligned}
	&h^2=\frac{B}{A}dr^2+\frac{C}{A}d\Omega^2,\\
	&W=\mathcal{U}(r)\frac{\partial}{\partial \phi},
\end{aligned}
\right.
\end{equation}
satisfying
		
	\begin{align}
		|W|^2=h_{33}(W^3)^2=\frac{C\sin^2\theta}{A}\mathcal{U}^2<1.
	\end{align}

We have
	
	\begin{align}
		&W_3=h_{33}W^3=\frac{C\sin^2\theta}{A}\mathcal{U},\nn\\
		&\lambda=1-|W|^2=1-\frac{C\sin^2\theta}{A}\mathcal{U}^2.\nn
	\end{align}
	
	The Randers metric~\eqref{RandersSSS} becomes
	
	\begin{equation}
		\begin{aligned}
			&\alpha^2=\frac{B}{A\lambda}dr^2+\frac{C}{A\lambda}d\Omega^2+\frac{C^2\mathcal{U}^2\sin^4\theta}{A^2\lambda^2}d\phi^2, \\
			& \beta=-\frac{C\mathcal{U}\sin^2\theta}{A\lambda}d\phi.
		\end{aligned}
	\end{equation}
	
  In this case, it is evident that $\beta$ is not closed. Consequently, the Randers metric does not feature reversible geodesics, and the NS metric corresponds to Eqs.~\eqref{Randerspacetime} or~\eqref{spacetime-zermole}. Utilizing $V^2=A\lambda=A-C\mathcal{U}^2\sin^2\theta$, the NS metric is given by
	
	\begin{align}
		\label{RTJW3821}
		g_{NS}=&-\left(A-C\mathcal{U}^2\sin^2\theta\right)dt^2-2C\mathcal{U}\sin^2\theta dtd\phi\nn\\
		&+B dr^2+Cd\Omega^2.	
	\end{align}

We apply the above navigation to flat spacetime and consider the following Zermelo pair with constant wind

\begin{equation}
	\left\{
		\begin{aligned}
			&h^2=dr^2+r^2d\Omega^2,\\
			&W=-\omega\frac{\partial}{\partial \phi}.
		\end{aligned}
		\right.
	\end{equation}
	
Here, $\omega$ is a constant satisfying $r^2\omega^2\sin^2\theta<1$. We immediately obtain the Langevin metric
	
		\begin{align}
		g_{NS}=&-\left(1-r^2\omega^2\sin^2\theta\right)dt^2+2r^2\omega\sin^2\theta dtd\phi\nn\\
		&+dr^2+Cd\Omega^2.	
	\end{align}
	
\subsection{Slowly rotating spacetime metric}

If we consider only the first-order contribution of $|\mathcal{U}|$, the NS metric~\eqref{RTJW3821} becomes

\begin{align}
	\label{SLMP}
	g_{NS}=&-Ad t^2-2C\mathcal{U}\sin^2\theta dtd\phi+B dr^2+Cd\Omega^2.
\end{align}

This metric is consistent with the metric of a slowly rotating SAS~\cite{Hartle-Thorne}. Therefore, it becomes reasonable to derive the metric for a slowly rotating spacetime by seeking specific wind on the optical space of the non-rotating static spacetime metric.

Interest in slow-rotating solutions arises from their suitability for various gravity tests. Additionally, certain gravity theories, such as the Einstein-bumblebee theory~\cite{DCSkerr} and the Einstein-aether theory~\cite{EAKerr}, have only identified slow-rotation solutions. The existence of exact rotating solutions remains an open question, emphasizing the importance of slow-rotating solutions in testing these theories.

Assuming that the wind has the following form

\begin{align}
\label{slowlySwind}
	W=\frac{\left(1-A\right)a}{C}\frac{\partial}{\partial \phi},
\end{align}
where $a$ is the angular momentum parameter, and

\begin{align}
	|W|^2=\frac{[(1-A)a\sin\theta]^2}{AC}<1.
\end{align}
Then, the NS metric~\eqref{SLMP} becomes

\begin{align}
	\label{SLMPLLL}
	g_{NS}=&-Ad t^2-2\left(1-A\right)a\sin^2\theta dtd\phi\nn\\
	&+B dr^2+Cd\Omega^2.
\end{align}
This metric describes a wide range of slowly rotating spacetimes.

\subsection{Mixed vector field}

Note that when $A=1$, it leads to $W=0$. Therefore, the slowly rotating spacetime cannot be obtained from flat spacetime using the Zermelo pair provided in Eq.~\eqref{ZP-rotating}.

According to {\it Theorem 2.5} in Ref.~\cite{Z-onRanders}, for a Riemannian metric $h$ and two winds $X$ and $Y$, the navigation of $(h,X+Y)$ and the navigation of $(F,Y)$ are equivalent, where $F=\alpha+\beta$ is obtained by navigating with $(h,X)$.

Connected with Sec.\eqref{RVF}, we can consider Euclidean space and mixed vector fields to form a Zermelo pair

\begin{equation}
	\left\{
	\begin{aligned}
		&h^2=dr^2+r^2d\Omega^2,\\
		&W=\mathcal{Q}\frac{\partial}{\partial r}+\frac{\mathcal{Q}^2a}{r^2}\frac{\partial}{\partial \phi},
	\end{aligned}
	\right.
\end{equation}
with

\begin{align}
	|W|^2=\mathcal{Q}^2+\frac{(\mathcal{Q}^2a\sin\theta)^2}{r^2}<1.
\end{align}

Then we have
\begin{align}
	&W_1=\mathcal{Q},\quad W_3=\mathcal{Q}^2a\sin^2\theta,\nn\\
	&\lambda=1-\mathcal{Q}^2+\frac{(\mathcal{Q}^2a\sin\theta)^2}{r^2} \approx 1-\mathcal{Q}^2,
\end{align}
where we only consider the linear terms of $a$.

Randers metric~\eqref{RandersSSS} is

\begin{equation}
	\label{MixRRdata}
	\begin{aligned}
		&\alpha^2=\frac{1}{1-\mathcal{Q}^2}\left(1+\frac{\mathcal{Q}^2}{1-\mathcal{Q}^2}\right)dr^2\\
		&~~~~~~~~+\frac{r^2}{1-\mathcal{Q}^2}d\Omega^2, \\
		& \beta=-\frac{\mathcal{Q}^2a\sin^2\theta}{1-\mathcal{Q}^2}d\phi+d\mathcal{Z}(r),
	\end{aligned}
\end{equation}
where

\begin{align}
	\mathcal{Z}(r)=-\int \frac{\mathcal{Q}(r)}{1-\mathcal{Q}(r)^2}dr.
\end{align}
Therefore, from the perspective of gauge invariance, we can remove $d\mathcal{Z}(r)$. Then Substituting Randers data~\eqref{MixRRdata} into Eq.~\eqref{Randerspacetime} and using $V^2=A\lambda=1-\mathcal{Q}^2$, we get the following NS metric

\begin{align}
	\label{EuZermeloSR}
	g_{NS}	=&-\left(1-\mathcal{Q}^2\right)dt^2-2\mathcal{Q}^2a\sin^2\theta\nn\\
	&+\frac{1}{1-\mathcal{Q}^2}dr^2+r^2d\Omega^2.
\end{align}

\subsection{Slowly rotating Kerr metric}

Starting with the Schwarzschild metric~\eqref{Schwarzschild}, we can construct the Zermelo pair $(h,W)$ by combining the optical metric and wind~\eqref{slowlySwind}, which can be expressed as follows

\begin{equation}
	\left\{
	\begin{aligned}
		&h^2=\frac{1}{\left(1-\frac{2M}{r}\right)^2}dr^2+\frac{r^2}{1-\frac{2M}{r}}d\Omega^2,\\
		&W=\frac{2Ma}{r^3}\frac{\partial}{\partial \phi},
	\end{aligned}
	\right.
\end{equation}
with

\begin{align}
	|W|^2=\frac{(2Ma\sin\theta)^2}{r^4(1-\frac{2M}{r})}<1.
\end{align}

As a result, this Zermelo pair transforms the NS metric~\eqref{SLMPLLL} into a slowly rotating Kerr metric, given by

\begin{align}
	\label{Kerr-metric}
	g_{K}=&-\left(1-\frac{2M}{r}\right)dt^2-\frac{2Ma\sin^2\theta}{r}dtd\phi\nn\\
	&+\frac{1}{1-\frac{2M}{r}}dr^2+r^2(d\theta^2+\sin^2\theta d\phi^2).
\end{align}

Now, consider the Zermelo pair formed by Euclidean space and mixed vector fields

\begin{equation}
	\label{FlatZerK}
	\left\{
	\begin{aligned}
		&h^2=dr^2+r^2d\Omega^2,\\
		&W=-\sqrt{\frac{2M}{r}}\frac{\partial}{\partial r}+\frac{2Ma}{r^3}\frac{\partial}{\partial \phi},
	\end{aligned}
	\right.
\end{equation}
with

\begin{align}
	|W|^2=\frac{2M}{r}+\frac{(2Ma\sin\theta)^2}{r^4}<1.
\end{align}

The NS metric~\eqref{EuZermeloSR} corresponding to Zermelo pair~\eqref{FlatZerK} is identical to Eq.~\eqref{Kerr-metric}.

\subsection{Slowly rotating Kerr-de Sitter metric}

Similarly, from the SdS metric~\eqref{SdS-metric}, we obtain the following Zermelo pair $(h,W)$

\begin{equation}
	\left\{
	\begin{aligned}
		&h^2=\frac{dr^2}{\left(1-\frac{2M}{r}-\frac{\Lambda}{3}r^2\right)}+\frac{r^2d\Omega^2}{(1-\frac{2M}{r}-\frac{\Lambda}{3}r^2)},\\
		&W= \frac{\left(2M+\frac{\Lambda}{3}r^3\right)a}{r^3}\frac{\partial}{\partial \phi},
	\end{aligned}
	\right.
\end{equation}
with

\begin{align}
|W|^2 = \frac{\left[\left(2M+\frac{\Lambda}{3}r^3\right)a\sin\theta\right]^2}{r^4\left(1-\frac{2M}{r}-\frac{\Lambda}{3}r^2\right)}<1.
\end{align}

Then, the slowly rotating KdS metric is given by

\begin{align}
	\label{Ebkds}
	g_{kds}=&-\left(1-\frac{2M}{r}-\frac{\Lambda}{3}r^2\right)dt^2\nn\\
	&-2\left(\frac{2M}{r}+\frac{\Lambda}{3}r^2\right)a\sin^2\theta dtd\phi\nn\\
	&+\frac{1}{1-\frac{2M}{r}-\frac{\Lambda}{3}r^2}dr^2+r^2d\Omega^2.
\end{align}

On the other hand, we use the following Zermelo pair
\begin{equation}
	\label{FlatZerKdS}
	\left\{
	\begin{aligned}
		&h^2=dr^2+r^2d\Omega^2,\\
		&W=-\sqrt{\frac{2M}{r}+\frac{\Lambda}{3}r^2}~\frac{\partial}{\partial r}+\frac{(2M+\frac{\Lambda}{3}r^3)a}{r^3}\frac{\partial}{\partial \phi},
	\end{aligned}
	\right.
\end{equation}
with

\begin{align}
	|W|^2 = \frac{2M}{r}+\frac{\Lambda}{3}r^2+\frac{[(2M+\frac{\Lambda}{3}r^3)a\sin\theta]^2}{r^4}<1.
\end{align}

Then, the NS metric~\eqref{EuZermeloSR} becomes KdS metric~\eqref{Ebkds}.

\section{Conclusion}
	\label{conclusion}
	
	Zermelo navigation is a fundamental tool in Finsler geometry. Moreover, in the field of physics, it serves as a powerful tool for geometrizing dynamics. By considering the Zermelo navigation problem in optical Riemannian space for static and spherically symmetric spacetime, we explore the process of generating new spacetime from existing spacetime using the Zermelo/Randers/spacetime triangle. Depending on whether the Randers metric, which is the solution of the Zermelo navigation problem, has reversible geodesics, we explore this problem in two scenarios, as illustrated in Fig.\ref{figLI}.
	
	In the first scenario, the Randers metric has reversible geodesics ($d\beta=0$), resulting in a static navigation spacetime. This condition is satisfied by a purely radial wind. In particular, we find that a variety of static spacetimes can be obtained from flat spacetime through Zermelo navigation. We provide specific examples, such as obtaining Schwarzschild spacetime through Newtonian wind and obtaining de Sitter spacetime through a cosmological constant wind. We also demonstrate that the same multi-parameter static spacetime can be obtained through different navigations. For instance, Schwarzschild-de Sitter spacetime can be derived from flat spacetime, Schwarzschild spacetime, and de Sitter spacetime, each achieved by considering different radial vector fields. However, for some spacetimes, certain navigation paths in real winds do not exist. For example, the RN spacetime can only be obtained from flat spacetime, not from Schwarzschild spacetime.
	
	The second route corresponds to the scenario where the Randers metric does not have reversible geodesics ($d\beta\neq0$). In this case, the navigation spacetime lacks time-reversal symmetry and is stationary. Our focus is on generating slowly rotating spacetimes. Starting from a non-flat static metric, we can choose a rotational vector field to achieve this goal. Starting from a flat metric, we can opt for a mixed vector field to achieve the same result. We illustrate both types of navigation with examples of Kerr spacetime and Kerr-de Sitter spacetime.
	
	The research presented in this paper is preliminary. We did not investigate the navigation problem within the context of Randers spaces, which would yield a new Randers metric~\cite{Z-onRanders}. Note that the optical metric of a stationary spacetime is a Randers metric. Therefore, by employing Zermelo navigation, it is possible to generate new stationary spacetimes from pre-existing ones. For example, one can obtain Kerr-AdS spacetimes starting from Kerr spacetimes. Furthermore, we anticipate that further investigations employing Zermelo navigation techniques may provide insights into various physical phenomena, including the equivalence principle, the Unruh effect, spacetime singularities, and more. We hope that by selecting appropriate wind vectors, we can derive complete stationary solutions from static solutions, akin to the Newman-Janis algorithm. Alternatively, Zermelo navigation may serve as a valuable tool for elucidating the Newman-Janis algorithm. The last interesting question, as suggested by the reviewer of this paper, is whether Zermelo navigation can be extended to obtain interior solutions. This concerns the validity of the entire framework below the horizon and should be investigated in future considerations.

	\acknowledgements
	This work is supported by grants from NNSF and MST China.

\end{document}